\documentstyle[11pt,aaspp4]{article}

\def\JN{$J_N=3_2-2_1$}
\def\nh3{NH$_3$}
\def\n2h{N$_2$H$^+$}
\def\c34s{C$^{34}$S}
\def\mm{$\mu$m }
\def\kms{km~s$^{-1}$}

\received{16 March 1999}
\accepted{12 April 1999}

\slugcomment{accepted to ApJL}

\lefthead{Ohashi et al.}
\righthead{Starless Core L1544 with Infall and Rotation}

\begin{document}

\title{CCS Imaging of the Starless Core L1544:\\
An Envelope with Infall and Rotation}

\author{Nagayoshi Ohashi and Siow Wang Lee}
\affil{Academia Sinica Institute of Astronomy \& Astrophysics\\
    P.O. Box 1-87, Nankang, Taipei 115, Taiwan; ohashi@asiaa.sinica.edu.tw}

\author{David J. Wilner}
\affil{Harvard-Smithonian Center for Astrophysics, MS 42\\
    60 Garden St., Cambridge, MA 02318}

\and

\author{Masahiko Hayashi}
\affil{Subaru Telescope, 650 North Aohoku Place, Hilo, HI 96720}



\begin{abstract}
We have carried out observations of the starless core L1544
in the CCS (\JN) line at 9 millimeters wavelength using the BIMA array. 
The maps show an elongated condensation, 
0.15 $\times$ 0.045~pc
in size, 
with stronger emission at the edges.
The appearance is consistent with a
flattened, ringlike structure viewed at high inclination to the line of sight.
The CCS molecule is likely heavily depleted in the inner part of the core.
The position velocity diagram along the
major axis shows a remarkable pattern, a \lq\lq tilted ellipse\rq\rq,
that can be  
reproduced by a simple model ring with motions of both infall and rotation.
The models suggest comparable velocities for infall 
and rotation, $\sim$0.1~\kms, in the outermost envelope, at radius 15000~AU.

\end{abstract}


\keywords{ISM: individual (L1544) --- ISM: kinematics and dynamics --- 
stars: formation}


%

\section{Introduction}

Radio and infrared observations have revealed that stars are formed 
in dense cloud cores (e.g., \cite{beich86,myers87}). 
From these observations as well as theoretical works, 
models of protostellar evolution for low-mass stars, 
from embedded protostars to visible T Tauri stars, 
have been proposed (e.g., Shu, Adams, \& Lizano 1987). 
Recently, even kinematic evidence for the infall of dense 
cores around embedded sources is observed (e.g., Zhou et al. 1993; 
Hayashi, Ohashi, \& Miyama 1993). Although our understanding of the embedded 
and T Tauri phases has made great progress, the earliest stage of 
star formation or even a prior stage to star formation has still 
been poorly understood. 

Dense cores without any detectable young stellar objects, 
i.e., starless dense cores are most probably sites where star 
formation has just started or will soon start. 
Although some dense cores, such as B 335 (Zhou et al. 1990), 
have been found to be a very early collapse phase, 
even in this case significant material has already collapsed 
onto the central star. Starless cores are therefore good targets 
to study the earliest stage of star formation. 
L1544 is a well-studied starless core in Taurus, observed in
several line emissions such as \nh3, CS, and \n2h 
(e.g., \cite{bm89,tafalla98}; hereafter T98, 
Williams et al. 1999; hereafter W99) and 
submillimeter continuum (Ward-Thompson et al. 1994).
Their observations were, however, made mainly using single-dish telescopes,
which prevented them from studying detailed geometrical and kinematical
structures of L1544.
In addition, most of the emission lines they used were optically thick, 
resulting in difficulty in 
investigating the whole velocity structure without suffering
self-absorption.
 
In this Letter, we report results of 
interferometric observations of the L1544 starless core in CCS (\JN).
This transition of CCS will be excited in cold cores such as
starless cores because of its low
energy level ($E_u\sim3.2$~K; \cite{yamamoto90}).
Interferometric observations enable us to study 
geometrical and kinematical structures of L1544 in detail.
In addition, CCS in L1544 is optically thin (see \S 2).
Moreover, CCS is one of the carbon chain molecules that are more abundant 
in starless cores (\cite{suzuki92}), 
and have no hyperfine structures, meaning that it is a good probe
to study geometrical and kinematical structures of starless cores in detail
(\cite{langer95}, Kuiper, Langer, \& Velusamy 1996, \cite{wolkovitch97}).

\section{Observations}

Observations were made using the nine-antenna 
Berkeley-Illinois-Maryland Association (BIMA) array
\footnote{Operated by the University of California at Berkeley, 
the University of Illinois, and the University of Maryland, 
with support from the National Science Foundation.}
in September 1998.
We observed CCS (\JN), which has rest frequency of 33.751374 GHz
(\cite{yamamoto90}), with
low-noise receivers utilizing cooled HEMT amplifiers, 
developed for observations of
the Sunyaev-Zeldovich effect (\cite{carl96}).
The filed of view of the array is $\sim$5\arcmin\ at this frequency. 
Such a wide field of view is of great advantage to the imaging of L1544, 
which is moderately extended (see T98).
The typical system temperature during 
the observations was 40-60~K in Single-side-band. 
Spectral information was obtained using a digital correlator 
with 1024 channels at a bandwidth of 6.25~MHz, 
providing a velocity resolution of $\sim$0.054 \kms\ 
at the observed frequency.
Two different configurations of the array were used. 
Projected baselines ranged from 6~m to 240~m, 
so that the observations were not sensitive to structures extended 
more than $\sim$5\arcmin, corresponding to 42000~AU at the distance of Taurus
($d=140$~pc; \cite{elias78})
\footnote{Note that the observations are less sensitive to the structures
close to 5\arcmin\ in extent (see Wilner \& Welch 1994)}. 
The obtained channel data was reduced using the MIRIAD package. 
The phase was calibrated by observing 0530+135, 
and the complex passband of each baseline was determined from 
observations of 3C84. When CCS maps were made and cleaned, 
a Gaussian taper was applied to the visibility data to improve
sensitivity to extended low brightness emission, 
so that the resultant beam size was 
20\arcsec$\times$13\arcsec\ with a position angle of $-5$\arcdeg. 
The resultant 1 $\sigma$ 
rms noise level for channel maps was typically 0.165~Jy~beam$^{-1}$, 
equivalent to $\sim$0.7~K in brightness.

In order to measure the optical depth of CCS in L1544,
we also observed CCS and CC$^{34}$S (\JN) simultaneously
using the Nobeyama 45~m telescope ($\Delta\theta\sim52\arcsec$) in April 1999.
$T_{\rm A}^*$ of CCS was measured to be 1.1~K, while no CC$^{34}$S emission
was detected to a 3 $\sigma$ level of 0.072~K in $T_{\rm A}^*$,
indicating that a 3 $\sigma$ upper limit to the optical depth of CCS
is 0.93 when the sulfur isotope ratio is the terrestrial value, 23.
In this Letter, we will discuss results of the BIMA observations in detail.




\section{Results}

CCS was detected at the LSR velocities
ranging from 6.88 to 7.48~\kms\ with high S/N ($\geq 3\sigma$).
Figure~1 shows the CCS total intensity map, integrated over
this velocity range. 
The map shows a structure elongated in the northwest-southeast 
direction (PA $\sim$144\arcdeg), with a size of 
$\sim$210\arcsec $\times$ 64\arcsec\ 
at the 3 $\sigma$ level, 
corresponding to 0.15 $\times$ 0.045~pc at the distance to Taurus.
The ratio between the major and minor axes is $\sim$3.3.  
The condensation consists of two blobs, one at the northwest and 
the other at the southeast, and each blob contains several peaks, 
suggesting that the condensation has clumpy structures.  


A similar elongated structure was also observed in \n2h (W99),
whereas it
shows a smaller size (7000~AU $\times$ 3000~AU) and
a slightly large position angle (PA$=$155\arcdeg).
More interestingly,
the CCS map does not show a prominent peak at the central part
of the condensation (marked by the cross in Fig.~1), 
where the other maps taken in \c34s (2-1), \n2h (1-0), and 800~\mm\ 
show prominent peaks (T98, W99, \cite{wt94}).
This difference suggests that the CCS in L1544
traces the outer regions of the core while the other molecules and 
dust trace the inner part.

The clumpy structures of the CCS condensation are more obvious 
in the channel maps as shown in the left panels of Fig.~2. 
Remarkable characteristics of the clumps are their 
narrow line width: each clump is detected at only a few velocity channels. 
This is shown in the right panels of Fig.~2, where
line profiles of two representative clumps are presented. 
The line widths deconvolved with the instrumental velocity resolution
(0.054~\kms) of the clumps 
were measured to be (0.06$\pm$0.01)-(0.13$\pm$0.01)~\kms\ in 
full width at half-maximum,
which are almost equivalent to the thermal line width of 
the CCS molecule at 10~K or 0.09~\kms\ (see Langer et al. 1995),
suggesting that thermal motions are dominant in the clumps.
Note that the line profiles of the clumps often show double peaks.
This is because
two clumps with different peak LSR velocities 
partially coincide with each other.
The multi-peak-velocity structure of L1544 was also observed in \c34s (T98).
Physical 
properties of the clumps will be discussed in detail in a forth coming paper.


We discern the global velocity field of the CCS condensation from
position-velocity diagrams (Fig. 3).  Two velocity components are
visible along the projected minor axis of the condensation
(Fig.~3a): one at 
$V_{\scriptsize LSR}\sim$7.1~\kms\ and the other 
at $V_{\scriptsize LSR}\sim$7.4~\kms.
Both components
persist all along the minor axis except at the southwestern edge, where
only the 7.4~\kms\ component is detected, and at the
northeastern edge, where only the 7.1~\kms\ component is detected. 
This suggests that the two
velocity components are mostly coincident along the line-of-sight.
Neither velocity component shows any significant gradient along the
minor axis.

The PV diagram along the projected major axis (Fig.~3b) shows more
remarkable velocity structures.
Similar to the PV diagram along the minor axis,
there are two velocity components in the inner parts of the condensation
($-40$\arcsec$ \leq \Delta x \leq +40$\arcsec\ in Fig.~3b),
whereas these two components merged into a single velocity component
at each of the southeastern and northwestern edges of the condensation.
The velocity difference between the two velocity components
at $\Delta x=0$\arcsec\ is $\sim$0.25~\kms, 
similar to that in Fig.~3a.
In addition, a global velocity gradient,
$\sim$0.08~\kms\ per 14000 AU, is observed from the southeast to the northwest.
The \n2h observations (W99) also showed a velocity gradient at inner parts
of L1544 along almost the same direction (PA$=$155\arcdeg)
although their measurements showed a $\sim$3 times larger gradient.
Thus, the whole velocity structure of the CCS condensation
can be represented as a
\lq\lq tilted ellipse\rq\rq, as shown by the dashed curve in Fig.~3b.
The elliptical velocity structure is
roughly symmetrical with respect to $V_{\scriptsize LSR}\sim$7.25~\kms,
suggesting that this velocity seems to be the systemic velocity of this system.
This suggests that the two velocity components seen in Fig.~3a
are actually blueshifted and redshifted parts 
in a single kinematical system.
The global velocity gradient seen along the projected major axis
suggests rotation of the CCS condensation, while the blueshifted and
redshifted components overlapping with each other along the line of sight
can be explained by inward motions in the condensation (see \S4.2.).


%
%

\section{Discussion}
\subsection{Geometrical Structures of the CCS Condensation}

The elongated structure seen in the CCS total intensity map suggests that 
the CCS condensation has either a filamentary structure or a flattened 
one with an almost edge-on configuration. 
The latter case is more likely 
in view of the high column densities needed to
produce the self-absorption features observed in optically thick 
molecular tracers such as CS (T98) or \n2h (W99).
The envelopes around young stellar objects (YSOs) 
often show 
flattened structures similar to the CCS condensation here
except that their sizes are smaller (e.g., \cite{ohashi97}).
This fact suggests that flattened envelopes
are present even before the YSOs form. 
The flattened geometry of starless cores are predicted by some theoretical 
simulations (e.g., \cite{nakamura95,matsumoto97}), in which
magnetic fields or rotation play an important role in producing
flattened structures.
Note that magnetic fields may be more important to explain the flattened
geometry of CCS because of its slow rotation (see \S4.2.; see also W99).
For the L1544 CCS condensation,
the ratio between 
the projected major and minor axes implies
an inclination angle of $\sim$73\arcdeg\ (0\arcdeg\ for the face-on case) 
if it is spatially thin. 
If the flattened condensation is spatially thick
like the envelopes around YSOs (e.g., L1527; \cite{ohashi97}),
then this estimate is a lower limit to the true inclination.
 
As pointed out in \S3, the CCS total intensity is 
stronger at the northwestern and southeastern edges of the 
condensation, and weaker in the center.
A ringlike geometry for the condensation, observed almost edge-on,
is consistent with this pattern, since, for optically thin emission, the 
total intensity is proportional to the column density
as long as there is no significant gradient in the excitation.
As we will show in the next section,
the velocity field also suggests a ringlike geometry for the condensation.
Note that the CCS ring is not a physical structure but rather probably 
results from a lower abundance of CCS towards the center of the L1544 core.
Most other tracers, including \c34s, \n2h, and submillimeter dust continuum,
show prominent peaks close to the center of the L1544 core. A similar
situation has been observed in L1498, another starless core in Taurus,
where CCS is weak toward the dust continuum peak (\cite{kuiper96,willacy98}).
The apparent structure may be due to chemical evolution.
As a dense core collapses and the density increases, the photoionization 
and photodissociation processes become gradually less effective in the
central region with highest density. The abundance of molecules sensitive
to the photochemistry changes, and the CCS abundance decreases substantially 
(\cite{suzuki92}). In addition, CCS depletes onto grains in a collapsing gas
that increases the density (\cite{bergin97}), 
also explaining the lower abundance of CCS toward the center of L1544. 
These ideas of the chemical evolution are consistent 
with the result that L1544
is undergoing infall, as evidenced by kinematic tracers (see \S4.2).

\subsection{Kinematical Structures of the CCS Condensation}

As shown in \S3, the kinematics of the CCS condensation are characterized
by two major features: (1) blueshifted and redshifted components that
coincide with each other along the line of sight, and (2) a velocity
gradient along the major axis. Taking into account the flattened edge-on 
structure of the CCS condensation, the blueshifted and redshifted components
may be explained by radial motion in the plane of the condensation,
while the velocity gradient may represent rotation of the condensation.
One possible radial motion in the plane of the condensation is infall,
which we favor over expansion, as discussed below. Here we consider the
kinematics of the condensation
using a simple model that consists of a spatially thin ring 
with both infall and rotation.  

A simple model ring was obtained by modifying the model disk used in
the study of the flattened envelope of L1527, which exhibits both infall
and rotation (\cite{ohashi97}).
For L1544, we have modified the 
surface density distribution and velocity field in two ways:
(1) The surface density is constant at 
$R_{\scriptsize out} > R > R_{\scriptsize in}$, and is null 
at $R \leq R_{\scriptsize in}$, 
where $R_{\scriptsize out}$ and $R_{\scriptsize in}$ are the outer and 
inner radii of the ring, 
respectively; (2) The infall velocity, $V_{\scriptsize infall}$, is constant 
throughout, while the rotation velocity, 
$V_{\scriptsize rotation}$, 
is proportional to the radius, i.e., rigid rotation such that 
$V_{\scriptsize rotation}=V_{\scriptsize rotation}^0$($R$/$R_{\scriptsize out}$), 
where $V_{\scriptsize rotation}^0$ is the rotation velocity 
at $R_{\scriptsize out}$. 
The first modification makes the model have a ring structure.
Under the assumption that the model ring has an edge-on configuration 
with respect to the observer, PV diagrams along the projected major 
axis of the model ring are calculated to compare with the observed 
PV diagram (Fig.~3a). 
For the model calculations, 
$R_{\scriptsize out}$ was fixed at 15000~AU according 
to the CCS observations, while 
three parameters, 
$R_{\scriptsize in}$, $V_{\scriptsize infall}$, 
and $V_{\scriptsize rotation}^0$ remained adjustable.

As shown in Fig.~4a, 
when $R_{\scriptsize in}=7500$~AU, 
$V_{\scriptsize infall}=0.12$~\kms, 
and $V_{\scriptsize rotation}^0=0.09$~\kms, 
the observed PV diagram (Fig. 3b) is well reproduced by the model: 
the calculated PV diagram shows a velocity structure, represented 
as a tilted ellipse, with a velocity difference at $\Delta x=0$ of
0.24~\kms\ and a global velocity gradient of 
0.09~\kms\ per 15000 AU. Note that $2V_{\scriptsize infall}$ 
corresponds to the velocity 
difference at $\Delta x=0$ in the 
calculated diagram, while $V_{\scriptsize rotation}^0$ is equivalent 
to the velocity 
gradient per 15000 AU in the diagram. Hence, it is easily 
understood that much smaller or larger $V_{\scriptsize infall}$ 
and/or $V_{\scriptsize rotation}^0$
cannot reproduce the observed velocity structures. On the other hand,
a ring structure with a larger $R_{\scriptsize in}$
is essential for the model to reproduce
the observed PV because
when
much smaller $R_{\scriptsize in}$, including the case of 
$R_{\scriptsize in}=0$ (equivalent to a model 
with a disk structure) is used,
prominent peaks emerge close to 
$\Delta x=0$ in the calculated diagram (see Fig. 4b). This is because when 
$R_{\scriptsize in}$ decreases, the total column density 
through the plane of the ring 
drastically increases close to $\Delta x=0$.

One might argue that expansion instead of infall can explain
the observed PV diagram. However, infall is more likely to explain
the kinematics of the CCS condensation because the mass of L1544
derived from the virial theorem
is comparable to or much smaller than
that estimated from the 800~\mm dust emission: 
a virial mass is estimated to be $\sim$1.7~$M_{\sun}$
using the radius of the CCS condensation (15000~AU) and the mean line
width of the CCS (0.4~\kms), under the assumption of a spherical cloud 
with a constant density,
while a mass of 2-6~$M_{\sun}$ was derived from 800~\mm dust emission
\footnote{Note that the current CCS data is not adequate to
derive a mass because CCS may be less abundant in the inner parts
of L1544 (see \S4.1).}
(T98).
Inward motions in L1544 were also suggested from
a spectroscopic method (T98, W99).
Our result is remarkable as the observations resolve the velocity field
and show {\it direct} evidence for infall motions
in the starless core L1544.

The kinematics of L1544 revealed by the CCS imaging show differences with
the envelopes of YSOs. The estimated infall and rotation velocities
are comparable to each other at a scale of 10000~AU, unlike YSOs where
infall velocities are inferred to be 2 to 6 times larger than rotation 
velocities in the outer regions. In addition, the estimated infall velocity
is much smaller in magnitude than those inferred around YSOs
(e.g., \cite{hayashi93,ohashi97,momose98}).

Although our model ring with a constant infall velocity and rigid 
rotation explain the kinematics of the CCS condensation very well, 
the current data do not place strong constraints on the radial 
dependences of the infall and rotation velocities because the CCS
molecule traces only the outer part of the L1544 core.
W99 have estimated infall and rotation velocities
in the inner regions of L1544, where CCS emission is weak or absent,
based on the self-absorption of \n2h.
Their estimates of
$\sim$0.08~\kms and $\leq$ 0.14~\kms for infall and rotation, respectively,
suggests nearly constant infall and rotation throughout the L1544 core,
from scales of 10000 to 1000~AU. 
The optical depth of the \n2h lines may yet mask the 
innermost regions of L1544, and direct measurements of the kinematics 
using an optically thin tracer may be valuable for revealing the
radial dependences of the systematic motions in this starless core.

\acknowledgments

We are grateful to P. T. P. Ho, H. Masunaga, F. Nakamura, K. Saigo, S. Takano, 
and S. Yamamoto for fruitful discussions. We also thank H. Maezawa for his
help during observations with the Nobeyama 45~m telescope.


\clearpage

\clearpage

\figcaption[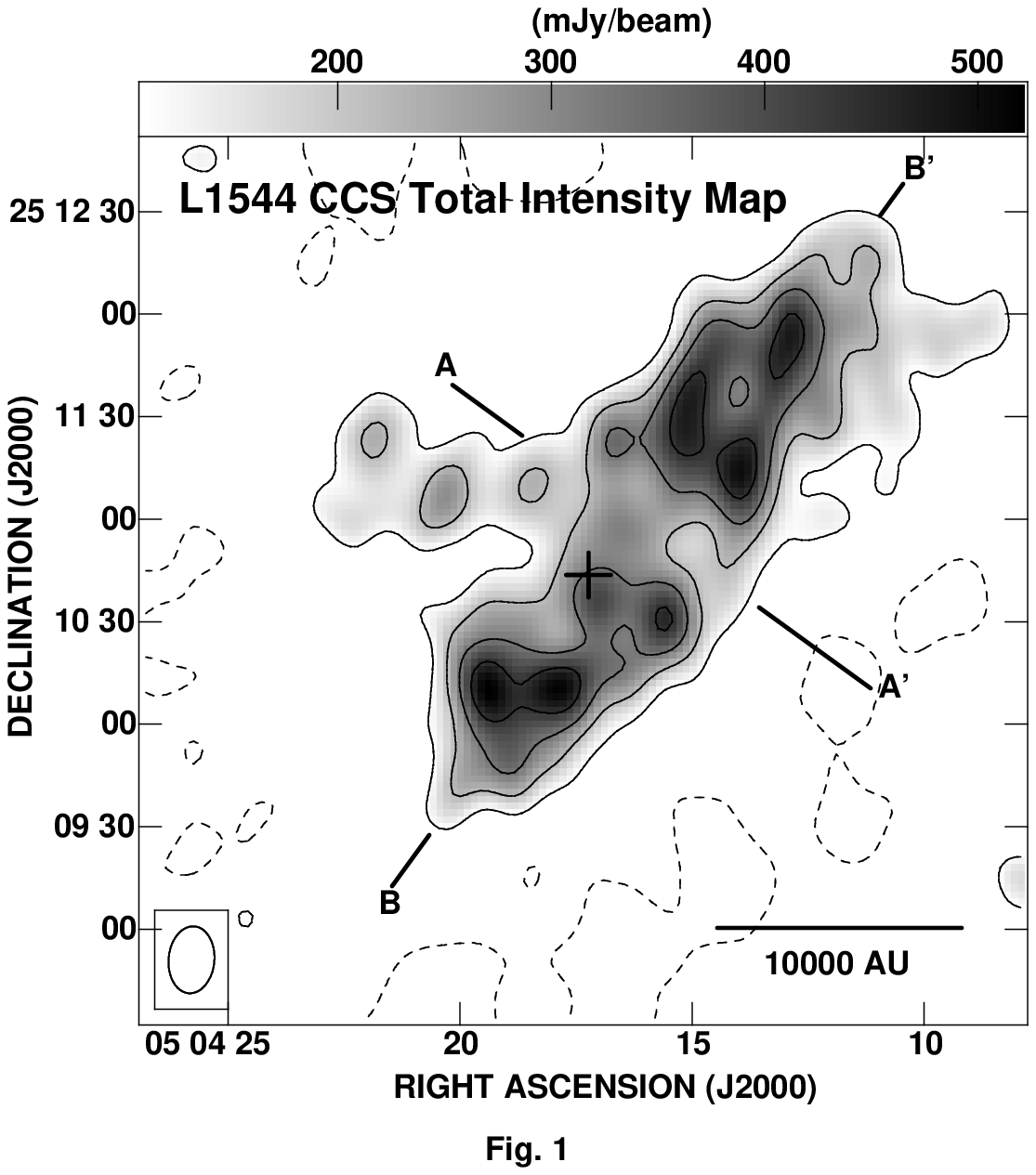]{The CCS total intensity map observed with
the BIMA array. The contour spacing is 2~$\sigma$, starting at $\pm 2~\sigma$
with $1~\sigma=55$~mJy~beam$^{-1}$. The cross indicates the peak position
of the 800~\mm continuum emission (Ward-Thompson et al. 1994). \label{fig1}}

\figcaption[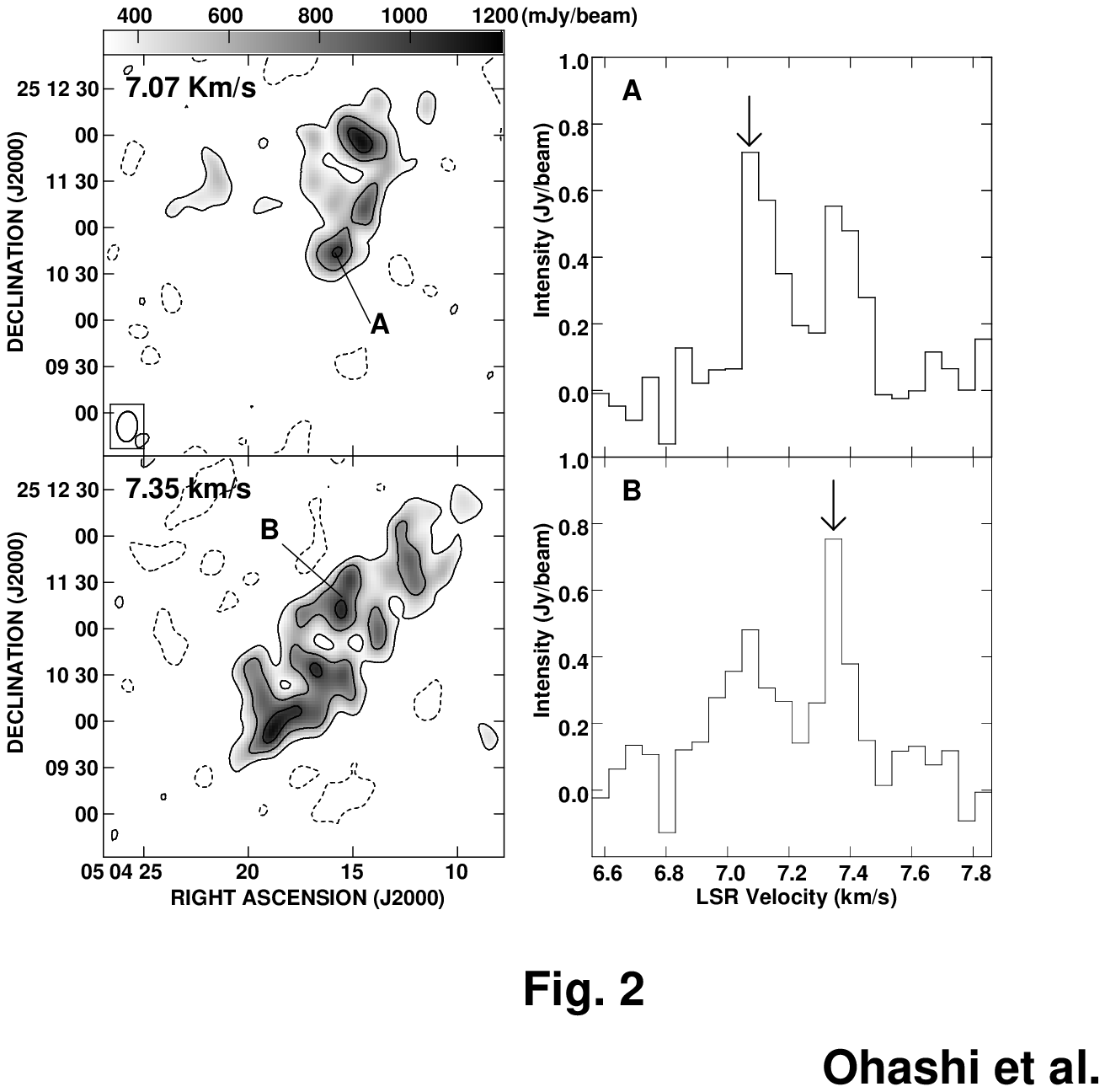]{({\it left panel}) Representative CCS channel maps.
The contour spacing is 2~$\sigma$, starting at $\pm 2~\sigma$
with $1~\sigma=165$~mJy~beam$^{-1}$.
{(\it right panel}) CCS line profiles of representative clumps (A and B)
identified in the channel maps shown in the left panel. A double velocity
component is obvious for each line profile. The arrows indicate
the corresponding velocity component for each clump. \label{fig2}}

\figcaption[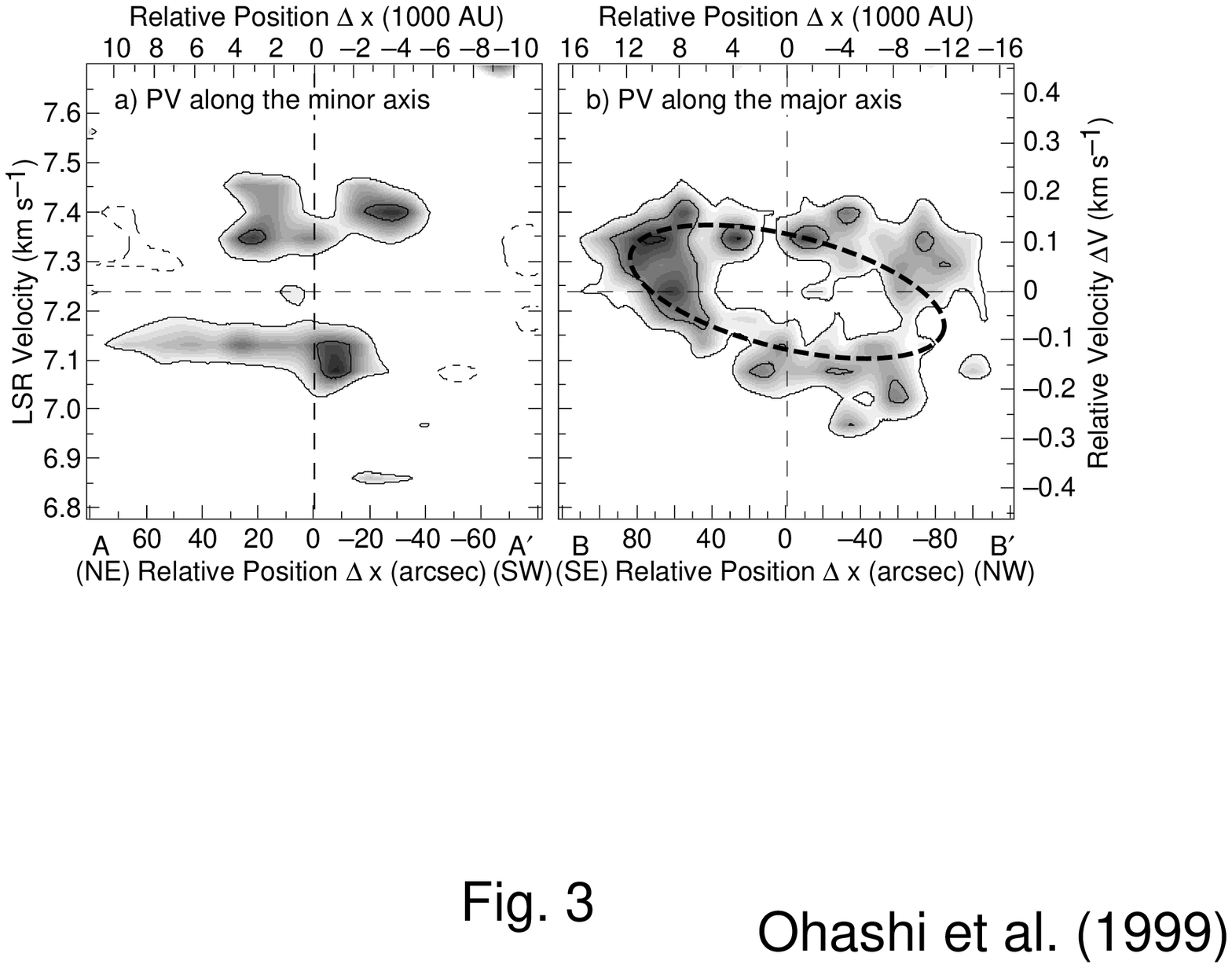]{The observed PV diagrams.
The contour spacing is 2~$\sigma$, starting at $\pm 2~\sigma$
with $1~\sigma=165$~mJy~beam$^{-1}$. 
The relative position ($\Delta x$) was measured
from [RA(2000)=5$^h$4$^m$15\fs86, Dec(2000)=25\arcdeg10\arcmin57\farcs6]. 
The velocity is presented in the LSR frame,
while the relative velocity ($\Delta V$) measured from 
$V_{\scriptsize LSR}=7.24$~\kms\
is presented in the righthand axis for comparison with the model calculations.
The two dashed lines indicate $\Delta x=0$ and $\Delta V=0$.
({\it a}) The PV diagram along the projected minor axis (PA$=$54\arcdeg;
line A-A' in Fig.~1).
({\it b}) The PV diagram along the projected major axis (PA$=$144\arcdeg;
line B-B' in Fig.~1). The dashed curve shows a tilted ellipse, which represents
the observed velocity structures. \label{fig3}}

\figcaption[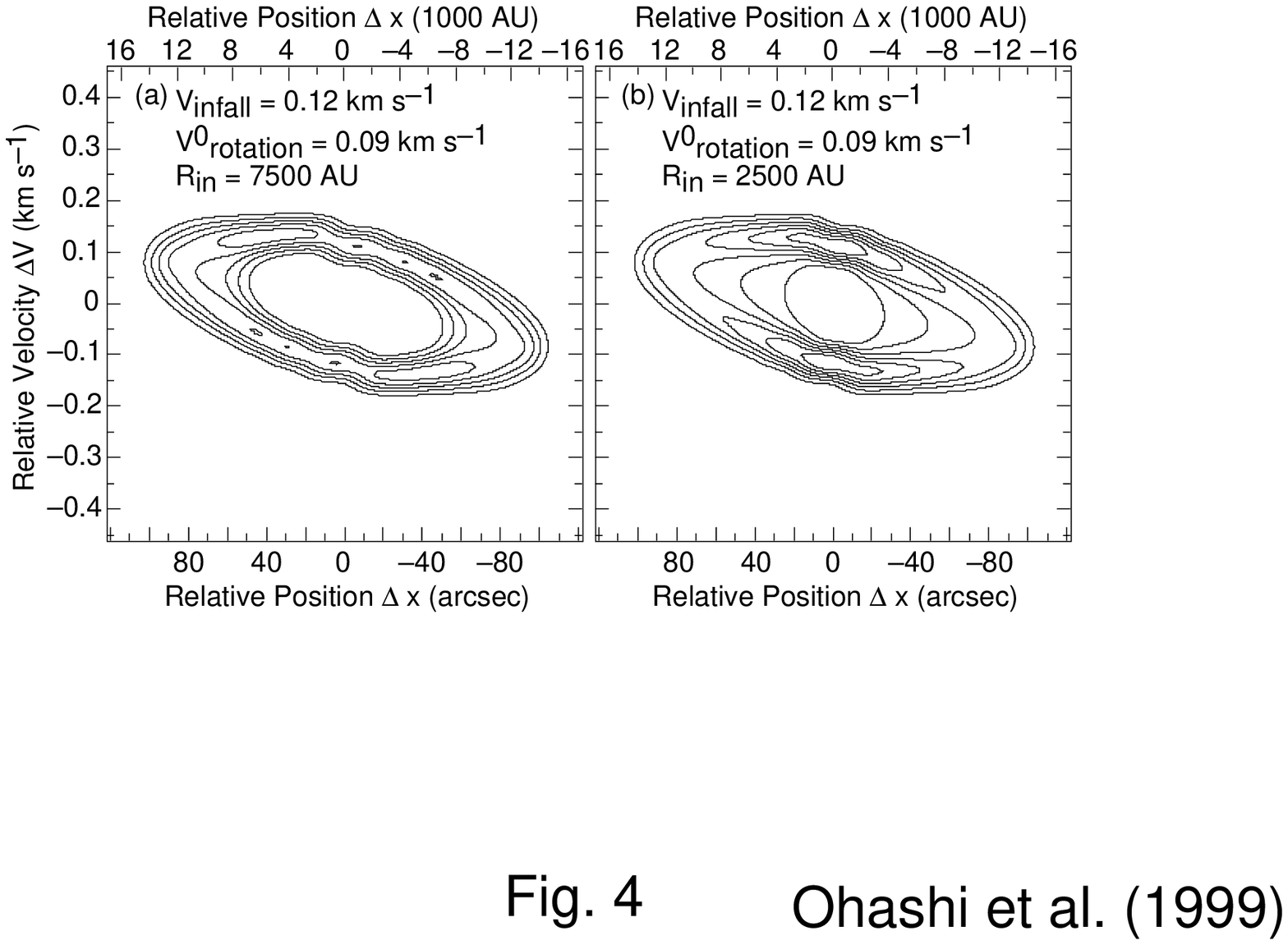]{The PV diagrams calculated two model rings.
({\it a}) The PV calculated with $V_{\scriptsize infall}=0.12$~\kms,
$V_{\scriptsize rotation}^0=0.09$~\kms, and $R_{\scriptsize in}=7500$~AU.
({\it b}) The same as (a) but $R_{\scriptsize in}=2500$~AU. \label{fig4}}



\clearpage

\plotone{1544_fig1.ps}

\clearpage

\plotone{1544_fig2.ps}

\clearpage

\plotone{1544_fig3.eps}

\clearpage

\plotone{1544_fig4.eps}

\end{document}